# Development of Novel Cathode Materials based on MWCNT for Energy Storage/Conversion Devices

Shruti Agnihotri, Anil Arya and Dr.A.L. Sharma*
*Centre for Physical Sciences, Central University of Punjab, Bathinda-151001, India*
*Corresponding Author E-mail: *alsharmaiitkgp@gmail.com*

*Abstract*

Novel compound comprising of $Li_2Fe_xMn_{(1-x)}SiO_4$/MWCNT were prepared by using Sol-Gel technique. In this study, improved electrochemical properties are observed by substituting Fe to replace Mn. and further multiwalled carbon nano tubes are used to prepare composite with prepared cathode material in Lithium ion battery and electrode materials in super capacitors. Various characterization techniques like: FESEM, EDX confirms the structure, morphology and texture of the material. Average particle size is found to be of order of 20-30 nm. As particles with smaller size reveal better electrical conductivity/electrochemical properties, which are confirmed by A.C Impedence spectroscopy and cyclic voltammetery results.

## I. INTRODUCTION

Electrodes are one of the essential components of the device where charge transfer reaction occurs via redox (oxidation-reduction) process. Selection of electrode materials, therefore, plays a crucial role in determining the device performance parameters. Double-layer charge storage in a supercapacitor is a surface process and hence surface characteristics of the electrodes greatly influence the capacitance of a supercapacitor. Similarly, the open circuit voltage, energy density, power density, cyclability and self life of a lithium polymer cell also depend on electrode material properties. In general, electrode should have the following properties; (i) high electronic conductivity (ii) high surface area (iii) high temperature stability (iv) controlled pore structure (v) low equivalent series resistance (ESR) and (vi) relatively low cost (cost effective) (vii) High electronic conductivity. Out of the aforementioned properties, foremost properties for improving the performance of Li ion battery are the development of suitable low cost, safety and high energy density cathode materials [1]. Recently, a new group of polyanion material $Li_2MSiO_4$ has been demonstrated as a promising candidate of Li ion insertion cathode material. Especially $Li_2MnSiO_4$ shows high theoretical capacity of 334 mAhg$^{-1}$. This is key feature of ortho silicate where two electron redox processes has been occurring. In $Li_2MSiO_4$ more than one Li ion per formula extraction is possible which significantly increases its experimental specific capacity and electrochemical performance.

In the present report, $Li_2Mn_xFe_{1-x}SiO_4$ is prepared using standard Sol Gel technique and the effect of iron (Fe) substitution on replacement of manganese (Mn) in $Li_2MnSiO_4$ of Fe for Mn [2] is seen. As polyanion materials $Li_2MSiO_4$ suffers from poor electronic conductivity. To improve the conductivity of oxide materials,further modification by making its composite with Carbon nano tube. As CNTs gained much attention due to their outstanding properties because of their unique structure. Field Emission scanning electron microscopy (FESEM) reveals the homogeneous topology of the materials sample. Impedance spectroscopy has been characterized for the estimation of electrical conductivity which is estimated of the order of ~ $10^{-6}$ S cm$^{-1}$.

## II. EXPERIMENTAL DETAILS

Sol-Gel technique is used to prepare $Li_2MnFeSiO_4$. Stoichiometric amount of $Li_2Co_3$, $MnO_2$, $Fe_2O_3$, $SiO_2$ are used as a starting materials. These precursors were hydrolyzed. Aqueous solution of above mentioned precursors were magnetically stirred. Further evaporation of solvent is done under continuous magnetic stirring at 80 °C. With continuous stirring and heating, a dry gel is formed. This dry gel is further ground to powder. Finally cathode material was sintered first heating at 350 ° C for 2 hrs and second heating 900° C for 12 hrs.

Synthesis of $Li_2Mn_xFe_{1-x}SiO_4$/MWCNT

Further to improve the particle –particle connectivity, composite of $Li_2MnFeSiO_4$/Multi walled carbon nano tube (MWCNT) was prepared via solution method. Stoichiometric amount of MWCNT was well dispersed in solution containing 10 ml distilled water and 10 ml acetonitrile through sonication.Then the active material (above prepared $Li_2MnFeSiO_4$)was added in sequence and finaly 0.5 ml of hydrazine was added. As obtained solution was refluxed at 80°c for two hrs with stirring to obtain $Li_2MnFeSiO_4$/MWCNT composite. This mentioned procedure was followed two times to prepare two samples with 6 wt percent of MWCNT and with 9wt. percent of MWCNT.These prepared samples are designated as LMFS-C6 and LMFS-C9 respectively.
. Morphology and particle size was observed by FESEM (Card Zeiss Merlin Compact ). Electrical conductivity and cyclic voltametry were performed by electrochemical analyzer (model: CH-760).

## III. RESULTS & DISCUSSIONS

From XRD analysis it is confirmed that cathode material exhibits orthorhombic structure with space group pmn21 [3]. FESEM is very powerful tool in order to get the surface morphology/topology in a very fine way. The high magnifications attainable combined with a large depth of field makes FESEM an outstanding diagnostic system for micro fabrication.

Figure 1 shows the Field emission electron microscopy (FESEM) image of the prepared cathode materials. The obtained image of the FESEM clearly depicts the homogeneous distribution of the matrials sample all over the sample.

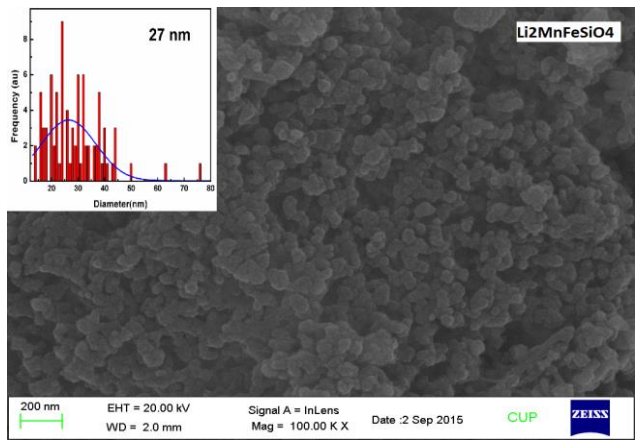

**FIG.1** FESEM analysis of $Li_2MnFeSiO_4$ (LMFS).

A small particle size could reduce the internal stress, because the binding force between particles may increase with the relative area of the grain boundaries. The particle size was measured from the software 'ImageJ' and the particle size was found to be of the order of ~27 nm (Fig. 1) for LMFS. This particle size is improved size as compared with $Li_2FeSiO_4$ and $Li_2MnSiO_4$. It is reported in literature that material with lesser particle size reveals better electrochemical properties due to reduction of path length of $Li^+$ ions [4]. Further particle size also decreases on addition of MWCNT.

The ac impedance measurement of the cathode material was carried out in the frequency range from 10 mHz to 1MHz at a input a.c. signal level of 10 mV. The impedance spectrum is comprised of a small semicircle at high frequency is followed by sharp spike at lower frequency (Fig 2).

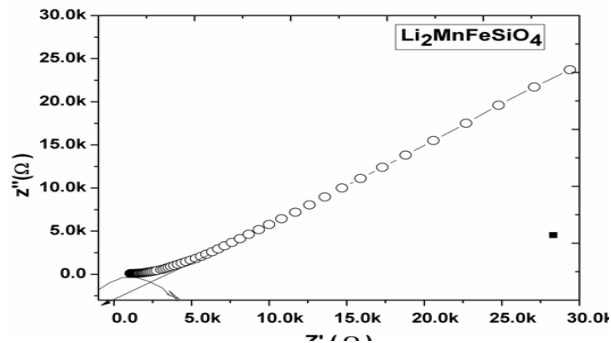

**FIG.2** Nyquist plot of $Li_2MnFeSiO_4$

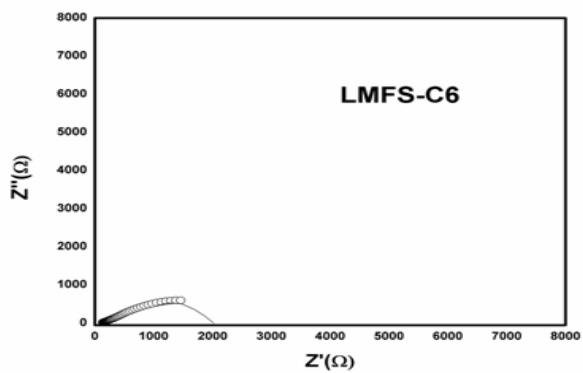

**FIG.3 Nyquist** plot of LMFS-C6

The small semicircle contribution is due the bulk response of the cathode materials whereas the lower frequency spike is due to the electrode electrolyte interface. The bulk resistance of the cathode materials is estimated by extrapolating the semicircle and cut on real axis, which comes out of order of ~2787ohm & ~2160ohm for LMFS & LMFS-C6 respectively (Fig 2&3). The lower frequency spike clearly shows the capacitive behavior of the cathode materials. The specific capacitance has been calculated for different scan rates using the formula; C= - (ω Z″)$^{-1}$ w+here ω (=2πf) is the angular frequency and Z″ is the imaginary part of the total complex impedance. The electrical conductivity of the cathode material is estimated by the formula: $\sigma_{dc} = \dfrac{1}{R_b}\dfrac{\ell}{A}$ . where, symbols have their usual meanings. The electrical conductivity has been calculated for LMFS &LMFS-C6 comes out to be of order of ~7.527× 10$^{-6}$ & 9.712× 10$^{-6}$ S cm$^{-1}$ respectively.

Cyclic Voltametery (CV) analysis was also done to study voltage stability window, energy density and specific capacity of the cathode materials. The recorded value of the electrochemical potential window is of the order of ~7 V. Specific capacity of prepared sample comes out of order of ~7.332μF.

## IV. CONCLUSIONS

In summary spherical Li$_2$MnFeSiO$_4$/MWCNT composite with excellent rate performance was successfully synthesized. Excellent electrochemical properties are caused by the unique composite structure. The nano sized Li$_2$MnFeSiO$_4$/MWCNT provide short path ways for Li$^+$ diffusion and CNT's network facilitates el$^-$ transport and avoids agglomeration of Li$_2$MnFeSiO$_4$. Our Results suggests that CNT play a profound part to increased efficiency of electrochemical properties. Prepared composite will have important application for energy storage and conversion in Li ion batteries.    .

## V. REFERENCES

.[1] Xiaozhen wu , Xinjhang ,Qisheng huo, Youxiang Zhang , *Electrochemica acta* , **80**, 50-55 (2012)
[2] BinShao, Y. L. Taniguchi, *Powder Technology* ,**235**, 1-8 (2013).
[3] Shivani Singh, Sagar mitra, *Electrochemica acta* **123**, 378-386 (2014)
[4] Zheng Zhang, Xingquan, Liping Wang, *Electrochimica acta*,**168**, 8-15, (2015).
[5] A. Rai, A.L. Sharma and A K Thakur, *Solid State Ionics*, **262**, 230-233 (2014).